\documentclass[letterpaper, 11pt]{article}
\pdfoutput = 1

\usepackage{shortcuts}

\usepackage[margin = 2.5cm]{geometry}

\usepackage{cuted}

\titleformat{\section}
  {\normalfont\fontsize{11}{13}\bfseries}{\thesection}{1em}{}

\begin{document}

\thispagestyle{empty}

\begin{flushright}
\texttt{BRX-TH-6661}
\end{flushright}

\begin{center}

\vspace*{5em}

{\LARGE \bf The Lattice-Continuum Correspondence \\ in the Ising Model}

\vspace{1cm}

{\large \DJ or\dj e Radi\v cevi\'c}
\vspace{1em}

{\it Martin Fisher School of Physics\\ Brandeis University, Waltham, MA 02453, USA}\\ \medskip
\texttt{djordje@brandeis.edu}\\

\vspace{0.08\textheight}
\begin{abstract}
{\normalsize Starting from the operator algebra of the $(1+1)$D Ising model on a spatial lattice, this paper explicitly constructs a subalgebra of smooth operators that are natural candidates for continuum fields in the scaling limit. At the critical value of the transverse field, these smooth operators are analytically shown to reproduce the operator product expansions found in the Ising conformal field theory.}
\end{abstract}
\end{center}

\newpage


\section{Introduction}

A fundamental property of any quantum field theory (QFT) is the structure of its operator algebra. This subject is ubiquitous: for example, it is the basis of an axiomatic approach to QFT \cite{Haag:1964}, it provides a storied phenomenological framework for particle physics \cite{GellMann:1962xb, Wilson:1969zs}, it underpins the conformal bootstrap \cite{Mack:1969rr, Polyakov:1974gs, Rattazzi:2008pe, ElShowk:2012ht}, it is studied under the guise of fusion categories in conformal and topological field theory \cite{Moore:1988qv, Moore:1989yh}, and it figures in defining various measures of entanglement \cite{Ohya:2004, Somma:2004, Casini:2013rba, Lin:2018bud}.

Precisely speaking, the preceding examples pertain to \emph{continuum} QFTs. Quantum theories on finite lattices are not usually associated to interesting algebraic structures. Their operators are finite matrices, and under usual matrix multiplication these form so-called type I algebras \cite{Murray:1936}. In contrast, operators of continuum QFTs are believed to form type III$_1$ algebras \cite{Halvorson2006}, with divergent operator traces and von Neumann entropies of subalgebras.

A sharp dichotomy between lattice and continuum theories cannot exist, however. Continuum QFTs provide excellent descriptions of systems with a large but finite number of degrees of freedom, particularly when such systems are near a second-order phase transition (see e.g.~\cite{Cardy:1996xt}). 
The operator-algebraic structure of the conformal field theory (CFT) describing this transition \emph{must} be encoded in the finite theory from the beginning.

It would thus be desirable to understand precisely how a continuum algebra of operators arises from a lattice one. We currently do not possess a universal method of establishing such a lattice-continuum correspondence. The state-of-the-art techniques study pairs of simple continuum and lattice theories in detail, and then manually match individual operators between them. The best studied example is the transverse field Ising model in $(1+1)$D, which can be matched to the Ising CFT at a critical value of the transverse field \cite{Kadanoff:1970kz, Koo:1993wz, Milsted:2017csn, Jones:2019, Zou:2019dnc}. Finding analogous correspondences for other minimal models is already rather nontrivial; analyses of the tricritical Ising and three-state Potts models have only been performed recently \cite{Mong:2014ova, OBrien:2017wmx, Zou:2019iwr}. See also \cite{Kitaev:2006lla} for a nice example in $(2+1)$D.

A direct, analytic approach to this issue was recently proposed by the present author. Starting from the usual algebra of lattice fermions in $(1+1)$D, Ref.~\cite{Radicevic:2019jfe} constructed the subalgebra generated by local lattice operators $\O(x)$ satisfying smoothness constraints of the form
\bel{\label{smoothness}
  \O(x + 1) = \O(x) + \hat\del \O(x),
}
with $\hat\del \O(x)$ having entries much smaller than $\O(x)$. Such smooth lattice operators were argued to behave precisely like continuum quantum fields. As one of several nontrivial checks of this proposal, the ``commutator'' of two natural operations --- projecting to this smooth algebra (``smoothing''), and taking a product of operators --- was shown to consistently reproduce the operator product expansion (OPE) of various quantum fields in the continuum fermion QFT.

The function of this paper is to use this smoothing method to analytically derive the lattice-continuum correspondence for the $(1 + 1)$D Ising model. To concretely illustrate how nontrivial operator-algebraic structures arise, continuum OPEs will be reproduced directly from the smoothed lattice operators. Previous studies have extracted OPE coefficients from the critical Ising system on the lattice by solving it and then comparing various correlation functions to expressions fixed by conformal symmetry (see \cite{Zou:2019dnc} and references therein). The present work will also rely on the solvability of the Ising model, but will not assume conformal (or any other) emergent symmetry when extracting the OPE coefficients. This will make the problem of finding noncritical OPEs transparent, and even tractable for a class of large perturbations away from the critical point. As a bonus, in all cases it will be possible to analytically establish the regime of validity of continuum QFT as a description of the lattice system.

\section{The critical Ising model}

The  Hamiltonian of the critical Ising model on a ring with $N$ sites is
\bel{\label{Ising}
  H = \sum_{w = 1}^N \left(Z_w + X_w X_{w + 1} \right), \quad Z_{N + 1} \equiv Z_1,
}
with $X$, $Y$, and $Z$ denoting the usual Pauli matrices. The Jordan-Wigner transformation maps the Ising spins to spinless Majorana fermions on $2N$ sites, via
\bel{
  \chi_{2w - 1} \equiv Z_1\cdots Z_{w - 1}X_w, \quad \chi_{2w} \equiv Z_1\cdots Z_{w-1} Y_w.
}
The fermions obey $\{\chi_v, \chi_u\} = 2\delta_{vu}$ for $1\leq v, u \leq 2N$. The theory \eqref{Ising} then becomes free,
\bel{
  H = - \sum_{v = 1}^{2N} \i \chi_v \chi_{v + 1}, \quad \chi_{2N + 1} \equiv - Q \chi_1,
}
where $Q \equiv \prod_{w = 1}^N Z_w$ generates the $\Z_2$ ``spin-flip'' symmetry of the Ising model, or the fermion number parity $(-1)^F$ in the dual fermion language.

This Hamiltonian can be diagonalized in each sector of the $\Z_2$ symmetry separately \cite{Schultz:1964fv}. Let
\bel{
  \chi_v \equiv \frac1{\sqrt{2N}} \sum_{k = -N}^{N - 1} \chi_k \,\e^{\frac{2\pi\i}{2N} kv} \equiv \frac1{\sqrt{2N}} \sum_{k = -N}^{N - 1} \~\chi_k \,\e^{\frac{2\pi\i}{2N} \left(k + \frac12\right) v}.
}
The transforms of $\chi_v$ satisfy $\{\chi_k, \chi_l\} = 2\delta_{k, -l}$ (with $\delta_{N, - N} \equiv 1$) and $\{\~\chi_k, \~\chi_l\} = 2\delta_{k, \, -l - 1}$. Thus positive and negative momentum operators behave as complex fermions and their conjugates. By letting $\psi_k\+ \equiv \frac1{\sqrt 2} \chi_k$ and $\~\psi_k \equiv \frac1{\sqrt 2} \~\chi_k$ for $k \geq 0$, the Hamiltonian can be written as
\bel{\label{Ising dual}
  H = \left\{
        \begin{array}{ll}\medskip
          -4 \sum_{k = 1}^{N - 1} \psi_k\+ \psi_k \sin\frac{2\pi k}{2N} + 2\cot\frac{\pi}{2N}, & Q = - 1:  \trm{ Ramond (R) sector}; \\ \smallskip
          -4 \sum_{k = 0}^{N - 1} \~\psi_k\+ \~\psi_k \sin \frac{2\pi (k +  1/2)}{2N} + 2 \csc\frac{\pi}{2N}, & Q = 1: \trm{ Neveu-Schwarz (NS) sector}.
        \end{array}
      \right.
}
The two ground states in the R sector both have $n_k \equiv \psi_k\+ \psi_k = 1$ for $0< k < N$, and the single ground state in the NS sector has $\~n_k = 1$ for $0 \leq k < N$. The NS ground state is the true vacuum. 

\section{The algebraic perspective}

The operator algebra $\A_N$ of the Ising model can be generated by any choice of $\{Z_w, X_w\}_{w = 1}^N$, $\{\chi_v\}_{v = 1}^{2N}$, $\{\chi_k\}_{k = - N}^{N - 1}$, or $\{\~\chi_k\}_{k = - N}^{N - 1}$. Integrating out all high-momentum modes turns \eqref{Ising dual} into a theory with a linear spectrum; in terms of operators, this is done by restricting to an algebra $\A_M$ generated by either $\{\chi_k, \chi_{k + N}\}_{k = - M/2}^{M/2}$ or by $\{\~\chi_k, \~\chi_{k + N}\}_{k = - M/2}^{M/2 - 1}$, with the effective lattice size $M \ll N$. (The choice of generating sets is immaterial in the scaling limit $M \gg 1$; this paper will focus on the former set.) Note that Majorana spinors $\Upsilon_k^\alpha \equiv (\chi_k, \chi_{k+ N})$, $\alpha \in \{+, -\}$, can be defined in analogy to the ``staggered'' Dirac fermions  \cite{Kogut:1974ag, Susskind:1976jm}.

Smoothing out the algebra $\A_M$ is achieved by projecting to its subalgebra $\A_M\^S$ generated by $\{\chi_k, \chi_{k + N}\}_{k = -k\_S}^{k\_S} \cup \{n_k, n_{N - k}\}_{k = k\_S + 1}^{M/2}$.  The cutoff $k\_S \ll M$ was called the ``string scale'' in Ref.~\cite{Radicevic:2019jfe} because it governs derivative expansions like eq.~\eqref{smoothness}; it can also be called the ``smoothness scale.'' The crucial point is that this second cutoff is \emph{necessary} to rigorously define the continuum limit of the lattice theory.

The occupation numbers $n_k^\alpha \equiv (n_k, n_{N - k})$ at $k > k\_S$ generate the center of $\A_M\^S$. In principle, each superselection sector, labeled by eigenvalues of these $n_k^\alpha$, corresponds to a different continuum theory. The sector of greatest interest is the one that contains the ground state. For the theory \eqref{Ising dual}, this is the sector labeled by $n_k^\pm = 1$ for all $k > k\_S$.

The position space Majoranas that generate $\A_M$ can be defined as
\bel{\label{chi xi}
  \chi_\xi \equiv \frac1{\sqrt{2(M + 1)}} \sum_{k = -M/2}^{M/2} \left( \chi_k \, \e^{\frac{2\pi\i}{2(M + 1)} k \xi} + \chi_{k + N} \, \e^{\frac{2\pi\i}{2(M + 1)} (k + M + 1) \xi} \right)
}
for $1 \leq \xi \leq 2(M + 1)$. They obey $\{\chi_\xi, \chi_\eta\} = 2\delta_{\xi\eta}$ and can be mapped to a new, coarser set of Ising spins via
\bel{\label{JW}
  \chi_{\xi = 2x - 1} \equiv Z_1 \cdots Z_{x - 1} X_x, \quad  \chi_{\xi = 2x} \equiv Z_1 \cdots Z_{x - 1} Y_x,
}
with $1 \leq x \leq M + 1$. The principal claim of this paper is that, after smoothing and some minimal processing, $Z_x$ and $X_x$ give rise to the familiar primary operators of the Ising CFT.

\section{Smoothing out $\Z_2$-even operators}

The smooth version of any operator $\O_{\vec x} \in \A_M$ will be denoted $\O(\vec x)$, with $\vec x$ possibly involving several lattice coordinates. For example, the Majorana operator $\chi_\xi$ from \eqref{chi xi} at $M \gg 1$ smoothes to
\bel{\label{chi(xi)}
  \chi(\xi) = \frac1{\sqrt{2M}} \sum_{k = -k\_S}^{k\_S} \left( \chi_k + (-1)^\xi \chi_{k + N} \right) \e^{\frac{2\pi\i}{2M} k \xi}.
}
Note that this operator obeys the smoothness relation $\chi(\xi + 2) = \chi(\xi) + O(k\_S/M)$.

Spin chain operators neutral under the global $\Z_2$ symmetry are the simplest to smooth out. In particular, it is straightforward to express $Z_x = -\i \chi_{2x - 1} \chi_{2x}$ in momentum space using eq.~\eqref{chi xi}, and then to project onto the algebra $\A_M\^S$ and obtain the smoothed operator $Z(x)$. The details are given in Appendix \ref{sec Wick}; the result is that $Z_x$ smoothes to the identity multiplied by $\avg{Z_x}$, namely
\bel{\label{Z(x)}
  Z(x) = - \frac 2\pi + O\left(\frac{k\_S}M \right).
}
The smoothness condition \eqref{smoothness} is obeyed as advertised (albeit trivially); this is ultimately because the contribution of modes above $k\_S$ is independent of $x$. Note that the algebra generated by $Z(x)$ for a fixed $x$ is very different from that generated by $Z_x$. For instance, while one simply has $Z_x^2 = \1$, powers of the smoothed operator $Z(x)^n$ show a nontrivial change in behavior at $n \sim k\_S$.

The operator $X_x X_{x + 1} = -\i \chi_{2x} \chi_{2x + 1}$ analogously smoothes to
\bel{
  XX(x, x + 1) = -\frac 2\pi + O\left(\frac{k\_S}M \right).
}
In terms of fermions, $Z_x$ and $X_x X_{x + 1}$ are formally related by $x \mapsto x + \frac12$. However, $XX(x, x + 1)$ is \emph{not} equal to $Z(x) + \frac12 \hat \del Z(x)$ when going beyond the leading order in $k\_S/M$. (Here $\hat \del$ is a formal derivative, with $\hat \del Z(x) = O(k\_S/M)$ because $\hat\del \e^{\frac{2\pi\i}{2M}k x} = \frac {\pi \i k} M \e^{\frac{2\pi\i}{2M}k x}$ and $|k| \leq k\_S$.) This is because smooth Majorana fields $\chi(\xi)$ and $\chi(\xi + 1)$ do not necessarily differ by an $O(k\_S/M)$ amount.

The operator
\bel{\label{Ex}
  E_x \equiv X_x X_{x + 1} - Z_x
}
captures this nontrivial distinction between $X_x X_{x + 1}$ and $Z_x$. Its smoothing, calculated in Appendix \ref{sec Wick}, is
\bel{
  E(x) = -2\i \chi(2x) \chi(2x + 1) + O\left(\frac{k\_S^2}{M^2} \right).
}
This operator is commonly associated with the energy density primary $\eps$ in the Ising CFT \cite{DiFrancesco:1997nk}. Without exploiting conformal symmetry, the OPE encoded by $\eps \times \eps = \1$ in the Ising CFT can be reproduced by studying $E_x$ and its smoothing, as shown next.

\section{Operator product expansions}

Intuitively, the OPE tells us about the short-range correlations that the lattice knows about but that the continuum operators themselves miss. Ref.~\cite{Radicevic:2019jfe} proposed to define the OPE of two operators $\O^{(1)}_x$ and $\O^{(2)}_y$ as
\bel{\label{OPE}
  \O^{(1)}_x \times \O^{(2)}_y \equiv \O^{(1)} \O^{(2)}(x, y) - \O^{(1)}(x)\O^{(2)}(y).
}
The r.h.s.~may be more familiar if written as $\O^{(1)} \O^{(2)} - \nord{\O^{(1)}\O^{(2)}}$, where the normal-ordering $\nord{\O}$ signals that the UV dependence of $\O$ was somehow removed. Thus eq.~\eqref{OPE} also serves  as a robust definition of normal-ordering.

The OPE $E_x \times E_y$ is readily computed by applying the procedure in Appendix \ref{sec Wick} to smooth out operators $E_x$ and $E_x E_y$. The result is
\bel{\label{OPE EE}
  E_x \times E_y = \frac{4/\pi^2}{(x - y)^2 - 1/4} + O\left(\frac{k\_S|x - y|}M\right).
}
Letting $\eps_x \equiv \frac\pi 2 E_x$ and taking $|x - y| \gg 1$ precisely gives the form expected from CFT, $\eps_x \times \eps_y = 1/(x - y)^2$. Note that even though $E(x) = O(k\_S/M)$, its OPE can still be an $O(1)$ operator.

This last observation can be clarified by the following discussion. The OPE \eqref{OPE EE} is valid only for separations $|x - y| \ll M/k\_S$, which gives an operational meaning to the usual statement that OPEs are sensible when two operators are ``close to each other.'' If $|x - y| \gtrsim M/k\_S$, the OPE vanishes, $E_x \times E_y = O(1/M)$. This means that the product of smoothed operators $E(x) E(y)$ captures correlations of microscopic operators $E_x$ and $E_y$ only when $|x - y| \gtrsim M/k\_S$. At shorter distances, $E(x) E(y)$ must be supplemented by the OPE data to get the microscopic correlations. The fact that $E(x)$ is small while its OPE is not means that the correlator of two $E_x$'s falls off fast enough, so that at $|x - y| \gtrsim M/k\_S$ one has $\avg{E_x E_y} = \avg{E(x) E(y)} = O(k\_S^2/M^2)$.

Note that $Z_x \times Z_y \approx 2/\pi^2(x - y)^2$, so $Z_x$ also obeys an OPE of the form $Z \times Z = \1$. Since $\avg{Z(x)} \neq 0$, $Z(x)$ cannot be interpreted as a scaling operator in a CFT: the identity is the only scaling operator with a nonzero expectation, while the OPE $Z \times Z$ indicates $Z$ has scaling dimension one. This is consistent with numerics finding that $Z_x$ is best approximated as a linear combination of $\1$, $\eps$, and their descendants in the Ising CFT \cite{Zou:2019dnc}.

\section{Smoothing out $\Z_2$-odd operators}

Spin chain operators charged under the $\Z_2$ symmetry, such as $X_x$, are more complicated to smooth. In the fermion picture, $X_x$ corresponds to the string
\bel{\label{Xx}
  X_x = (-\i)^{x - 1} \chi_1 \cdots \chi_{2x - 1}.
}
Even though each Majorana has a simple smoothing \eqref{chi(xi)} on its own, a macroscopically large number of fermions is harder to work with. Appendix \ref{sec strings} shows that $X(x)$, at least for $x \ll M$, can be expressed as a sum of smoothed Majorana operators
\bel{\label{X(x)}
  X(x) = (-1)^{x-1} \sum_{x' = 1}^x b_{x'}\, \chi(2x' - 1),
}
with $b_{x'}$ being $O(1)$ coefficients that can be determined by evaluating certain Toeplitz determinants and minors. Extending this result to $x \sim M$ would only be possible if operators like $\chi(x_1) \cdots \chi(x_m)$, with $m \sim M$, could be neglected in the smoothing. However, a quick way to see that this is \emph{not} so is to consider the operator $X_{M + 1}$, which simply smoothes to $X(M + 1) = \i Q \chi(2M + 2)$, where $Q$ is the generator of the $\Z_2$ symmetry. This smoothing is not compatible with a naive extrapolation from eq.~\eqref{X(x)}, which means that finite-$M$ effects will necessarily become important at $x \sim M$.

A much more reliable calculation can be done to find the OPE $X_x \times X_y$. For $|x - y| \ll M/k\_S$, which is the usual regime of interest for OPEs as discussed below eq.~\eqref{OPE EE}, Appendix \ref{sec strings} finds
\bel{\label{OPE XX}
  X_x \times X_y =  \frac{(-1)^{x - y} c_0}{|x - y|^{1/4}} \left( \1 - \frac 12  |x - y|\, \eps(x)\right) + O\left(\frac{k\_S |x- y|}{M} \right),
}
with $c_0 = 0.6450(1)$. (The operator $\eps_x$ was defined below eq.~\eqref{OPE EE}.) All the numerical prefactors and exponents shown above were determined to three or more significant digits in the limit $|x - y| \gg 1$.

It is thus possible to define
\bel{
  S_x \equiv \frac{(-1)^{x - 1}}{\sqrt{c_0}}  X_x
}
and to notice that its smoothed version $S(x)$ behaves precisely as the spin operator $\sigma$ in the Ising CFT, having a vanishing vacuum expectation and satisfying the schematic OPE  $\sigma \times \sigma = \1 + \eps$ with the correct OPE coefficient $C_{\sigma\sigma\eps} = -\frac12$ \cite{DiFrancesco:1997nk}. This is consistent with the recent numerical results identifying the lattice operator $X_x$ with the primary $\sigma$ multiplied by $\sqrt{c_0} \approx 0.803$ \cite{Zou:2019dnc}.

\section{OPEs away from the critical point}

The analysis so far naturally extends to perturbations that destroy criticality. There are two classes of deformations of the critical Hamiltonian $H$ defined in \eqref{Ising}.

In the first class are deformations by operators in $\A_M\^S$. By definition, such deformations all commute with the center of $\A_M\^S$, but they may change the superselection sector in which the ground state resides. For instance, the ground state of $H + \lambda \sum_x E(x)$ lies in the same sector as the ground state of $H$, namely the one labeled by $n_k^\alpha = 1$. On the other hand, the ground state of $H + \mu \sum_k n_k$ lies in a different sector if $\mu$ is large enough. The coupling $\mu$ acts as a chemical potential in the free fermion theory. In the limit $\mu \rar \infty$, the ground state has $n_k^\alpha = 0$.

The second class of deformations involves operators outside $\A_M\^S$. These  will generically fail to commute with $n_k^\alpha$. If we wish to think of the structure of an operator algebra as constant in time, we will need to change the smoothing procedure. For an illuminating example, consider the lattice theory
\bel{\label{Ising h}
  H_h = \sum_{w = 1}^N (X_w X_{w + 1} + h Z_w)
}
for $h \neq 1$. This theory is still exactly solvable. It bosonizes to a massive free fermion, whose momentum modes $\varphi_k$ in the R sector are related to the fermions from eq.~\eqref{Ising dual} via
\bel{
  \psi_k \equiv u_k \varphi_k + \i v_k \varphi_{N - k}\+
}
for momenta $1 \leq k < \frac M 2$ and for $h$-dependent Bogolyubov coefficients $u_k$, $v_k$ \cite{Schultz:1964fv}. The natural smoothing procedure would project to an algebra $\hat\A_M\^S$ whose sole generators above $k\_S$ are
\bel{
  \hat n_k \equiv \varphi_k\+ \varphi_k = u_k^2 n_k + v_k^2 n_{N - k} + \i u_k v_k \left( \psi_k\+ \psi_{N - k}\+ + \psi_k \psi_{N - k}  \right).
}
The center of this algebra has significant overlap with the center of the original algebra $\A_M\^S$: operators of the form $(-1)^{n_k + n_{N - k}}$ belong to both centers. It is thus possible to systematically interpolate between the algebras for each momentum $k$ separately. A generic perturbation of the microscopic Hamiltonian \eqref{Ising} would not be so tractable.

The change of the superselection sector, or of the center itself, typically leads to a different algebraic structure in the continuum. For instance, a theory with a large chemical potential can have $n^\alpha_k = 1$ only for $\frac M 2 < k < M$. This theory would have significantly different OPE coefficients. In particular, the methods of Appendix \ref{sec strings} can be applied to show that the OPE $X_x \times X_y$ still has the schematic form $X \times X = \1 + E$ as in \eqref{OPE XX}, but this time the coefficients decay as $1/2^{|x - y|}$. The characteristic length in this exponential falloff is set by the filling factor of modes above $k\_S$. An analogous exponential decay is obtained for the model \eqref{Ising h} in a sufficiently large external field $h$.

\section{Outlook}

The OPEs \eqref{OPE EE} and \eqref{OPE XX} are the main results of this paper. Their agreement with OPEs of the Ising CFT further supports the proposal of Ref.~\cite{Radicevic:2019jfe} that smoothed lattice operators directly correspond to the familiar continuum fields. Unlike the numerical approaches of Refs.~\cite{Mong:2014ova, OBrien:2017wmx, Milsted:2017csn, Zou:2019dnc, Zou:2019iwr}, the smoothing approach is not limited to identifying lattice versions of primaries and their immediate descendants, and instead it constructs the lattice-continuum correspondence for the entire algebra, with precision controlled at will by the ``string scale'' $k\_S$. Indeed, this is what allowed Ref.~\cite{Radicevic:2019jfe} to formulate high-momentum corrections to Abelian bosonization.

Although the methods of this paper are limited to theories of fermions, they may in principle be used to understand the lattice-continuum correspondence in any spin chain, thanks to the Jordan-Wigner transformation. The calculations in this work crucially depended on the fact that the relevant fermion theories were free, but the overall method of smoothing does not require this assumption. The methods given here can be used to calculate OPEs in any noncritical model with only smooth (``first-class'') deformations away from criticality. It would be interesting to see whether less pedestrian (potentially numerical) methods can be developed to handle the smoothing of interacting lattice fermions with nonsmooth (``second-class'') deformations. In particular, it is important to understand how the algebraic structure can be extracted from a smoothed algebra in which the superselection sector labels $n_k^\alpha$ change with time.

Many other topics for future work were listed in Ref.~\cite{Radicevic:2019jfe}. One topic, however, is so immediate in its relevance that it deserves special mention here: the explicit identification of Virasoro generators as elements of $\A_M\^S$. All questions regarding the conformal symmetry and radial quantization form a logical unit whose analysis will be published in a separate publication.

\section*{Acknowledgments}

It is a pleasure to thank Ruben Verresen for a useful discussion and Nick Jones for pointing out additional references and for help with fixing the minus signs. This work was completed with the support from the Simons Foundation through \emph{It from Qubit: Simons Collaboration on Quantum Fields, Gravity, and Information}, and from the Department of Energy Office of High-Energy Physics through Award DE-SC0009987.

\appendix

\section{Smoothing out and Wick contractions} \label{sec Wick}

There is a systematic way to smooth out operators of the form
\bel{\label{app example}
  \frac1{M^{(n + m)/2}} \sum_{k_i, \, k'_i = -M/2}^{M/2} \chi_{k_1} \cdots \chi_{k_n} \chi_{k'_1 + N} \cdots \chi_{k'_m + N}  \, \e^{\frac{2\pi \i}{2M} \left(\vec k \cdot \vec \xi + \vec k' \cdot \vec \xi'\right)}.
}
The key fact is that, among the terms in which some momenta have $|k_i| > k\_S$, only terms in which two $\chi$ operators combine into the identity or into $n_k^\alpha$ will survive the smoothing.

Contractions of two Majoranas into the identity will always cancel out when smoothing local operators, i.e.~when summing all terms of the form \eqref{app example} that figure in a Fourier expansion of products of several $\chi_\xi$'s. These contractions are not interesting and will be denoted by an ellipsis (\ldots).

More interesting are terms in the sum \eqref{app example} in which two $\chi$'s combine into a momentum occupation number $n^\alpha_k$. This happens when their momenta are related by $k_i = - k_j$. When $|k_i| > k\_S$ in this situation, the $n_{k_i}^\alpha$ are c-numbers. It is convenient to think of such terms as Wick contractions familiar from QFT. (Indeed, this will provide a precise understanding of the sense in which OPEs arise from integrating out high-momentum modes, as advocated in \cite{Weinberg:1996kr}: these modes are not integrated out in a Wilsonian sense, but instead they are classical variables that are summed to give OPE coefficients.) The rules for Wick contractions are
\gathl{\label{rules Wick}
  \wick[offset = 1.6ex]{\c\chi_{k_i} \c\chi_{k_j}} = 2 \delta_{k_i, - k_j} \, \theta(k_i - k\_S), \\
  \wick[offset = 1.6ex]{\c\chi_{k_i} \c\chi_{k_j + N}} = \wick[offset = 1.6ex]{\c\chi_{k_i + N} \c\chi_{k_j}} = 0, \\
  \wick[offset = 1.6ex]{\c\chi_{k_i + N} \c\chi_{k_j + N}} = 2\delta_{k_i, - k_j}\, \theta(k_j - k\_S).
}
The step functions (with $\theta(0) = 1$) are present because $\chi_k$ and $\chi_{N - k}$ for $k > 0$ are interpreted as creation operators. If, say, $k_i < 0$ in the contraction of $\chi_{k_i}$ and $\chi_{k_j}$, the resulting $c$-number would be $1 - n_{k_i}^+ = 0$, and so such contractions can be disregarded.

The smoothing of \eqref{app example} is thus given by the sum over all possible contractions, with the proviso that all uncontracted fermions have momenta $k_i$ or $k_i'$ running only between $-k\_S$ and $k\_S$. The sum over momenta greater than $k\_S$ is given by
\bel{\label{sum Wick}
  \frac1M \sum_{k_i = k\_S + 1}^{M/2} \e^{\frac{2\pi\i}{2M} k_i (\xi_i - \xi_j) } = \frac{\i}{\pi} \frac{1 - \e^{\i \pi (\xi_i - \xi_j)/2}}{\xi_i - \xi_j} + O\left(\frac{k\_S|\xi_i - \xi_j|}M \right).
}
If $\xi_i = \xi_j$, the sum is exactly $1/2$. If $|\xi_i - \xi_j| \gtrsim M/k\_S$, the summand becomes a rapidly oscillating function, and the total sum becomes $O(1/M)$. The result \eqref{sum Wick} is clearly related to the position-space propagator of a free fermion. However, it is important to keep in mind that this connection is only possible because the high-momentum operators $n_k^\alpha \in \A_M\^S$ are quadratic in the original fields.

As an example, consider the smoothing out of the operator $Z_x = -\i \chi_{2x - 1} \chi_{2x}$. By eq.~\eqref{chi xi}, to leading order in $k\_S/M$ this operator is given by
\bel{\label{Zx}
  Z_x  = \frac{-\i}{2M} \sum_{k, \, l = -M/2}^{M/2}\!\! \left(\chi_k \chi_l + \chi_k \chi_{l + N} - \chi_{k + N} \chi_l - \chi_{k + N} \chi_{l + N} \right) \e^{\frac{2\pi\i}{2M} [(2x - 1) k + 2x l]}.
}
Only the first and the fourth term will have nontrivial Wick contractions; in the other two terms, it is sufficient to simply restrict the momentum sum to $|k| \leq k\_S$. The $\chi_k \chi_l$ term in \eqref{Zx} smoothes to
\bel{\label{temp sum}
  -\frac{\i}{2M} \sum_{k, \, l = -k\_S}^{k\_S} \!\! \chi_k \chi_l \, \e^{\frac{2\pi\i}{2M} [(2x - 1) k + 2x l]} - \frac\i M \sum_{k = k\_S + 1}^{M/2}\!  \wick[offset = 1.6ex]{\c \chi_k \c \chi_l} \, \e^{-\frac{2\pi\i}{2M} k} + \ldots
}
The first sum above will combine with the corresponding uncontracted sums from the other three terms in \eqref{Zx} to give a product of smoothed fermion fields. The second sum can be calculated using eq.~\eqref{sum Wick}, giving
\bel{
  - \frac {1+ \i}\pi + \ldots + O\left(\frac{k\_S}M \right).
}
The contraction in the $\chi_{k + N}\chi_{l + N}$ term in \eqref{Zx} is evaluated the same way, giving $-(1 - \i)/\pi + \ldots + O(k\_S/M)$. Adding this up, the ellipsis terms cancel out, and the result is
\bel{
  Z(x) = -\frac 2\pi - \i \chi(2x - 1) \chi(2x) + O\left( \frac{k\_S}M \right).
}

The product $\chi(2x - 1) \chi(2x)$ is an operator whose entries are all $O(k\_S/M)$. To see this, consider the first sum in eq.~\eqref{temp sum}; call it $S$. Now, relabel $k \leftrightarrow l$ in $S$, and then exchange the positions of the Majoranas. This manipulation produces the relation $S = -S + O(k\_S/M)$, which means that $S$ is indeed small. The same argument will apply to the sums involving $\chi_{k + N}\chi_{l + N}$ and $\chi_{k + N}\chi_l - \chi_k \chi_{l + N}$. Therefore it is possible to write
\bel{
  Z(x) = -\frac2\pi + O\left(\frac{k\_S}M \right),
}
as advertised in eq.~\eqref{Z(x)}.

When smoothing the operator $E_x$, defined in eq.~\eqref{Ex}, all the contractions cancel. The leading nontrivial term in the result is $O(k\_S/M)$ and is given by
\bel{
  E(x) = -\i \left[\chi(2x) \chi(2x + 1) - \chi(2x - 1)\chi(2x)\right].
}
Recall that $\chi(\xi + 1) \neq \chi(\xi)$ even to leading order in $k\_S/M$, so the two terms in brackets do not cancel at leading order. However, since $\chi(\xi + 2) = \chi(\xi) + O(k\_S/M)$, it is possible to write
\bel{
  E(x) = -2\i \chi(2x) \chi(2x + 1) + O\left(\frac{k\_S^2}{M^2} \right).
}
Finally, starting from momentum-space Majorana spinors $\Upsilon^\alpha_k = (\chi_k, \chi_{N + k})$, it is possible to first define position-space spinors $\Upsilon^\alpha(x)$ and then to show that $E(x)$ corresponds to
\bel{
  E(x) = 2\i \, \Upsilon^+(x) \Upsilon^-(x).
}
This is, up to normalization, the definition of the $\eps$ primary presented in Ref.~\cite{DiFrancesco:1997nk}.

\section{Smoothing strings of operators} \label{sec strings}

The goal of this section is to understand the smoothing of strings of fermions $\chi_{\xi_1} \ldots \chi_{\xi_n}$ using the Wick contractions developed in Appendix \ref{sec Wick}. The first step is to develop position-space Wick contractions. At $M \gg 1$, the Fourier transform \eqref{chi xi} can be written as
\bel{
  \chi_\xi = \frac1{\sqrt{2M}} \sum_{k = -M/2}^{M/2} \left( \chi_k + (-1)^\xi \chi_{k + N}  \right) \e^{\frac{2\pi\i}{2M} k \xi}.
}
Recall that contracting two fermions really means isolating those terms in which their momenta obey $k = - l$. If these momentum modes come from the Fourier expansion of $\chi_\xi \chi_\eta$, there will be two possible contractions associated to coordinates $\xi$ and $\eta$, and they can be captured by
\bel{
  \wick[offset = 1.6ex]{\c \chi_\xi \c \chi_\eta} \equiv \frac1{2M} \sum_{k,\, l = -M/2}^{M/2} \left( \wick[offset = 1.6ex]{\c \chi_k \c\chi_l} + (-1)^{\xi - \eta} \wick[offset = 1.6ex]{\c \chi_{k + N} \c\chi_{l + N}} \right) \e^{\frac{2\pi\i}{2M} (k \xi + l \eta)}.
}
Using the rules \eqref{rules Wick} and the sum \eqref{sum Wick}, this expression can be rewritten as
\bel{\label{propagator Wick}
  \wick[offset = 1.6ex]{\c \chi_\xi \c \chi_\eta} =
  \frac{\i}\pi \frac {1 - (-1)^{\xi - \eta}}{\xi - \eta} +  O\left(\frac{k\_S|\xi - \eta|}M \right).
}
This is the short-distance correlator of two free fermions, consistent with the result \eqref{Z(x)} for $Z(x)$.

Now consider how this works for $X_x$, given by eq.~\eqref{Xx}. The aim is to sum over all possible Wick contractions. For small $x$, this is a simple task. For instance, for $x = 1$ there are no contractions and the result is
\bel{
  X(1) = \chi(1),
}
with $\chi(\xi)$ defined in eq.~\eqref{chi(xi)}. For $x = 2$, the sum over the two possible contractions gives
\bel{
  X(2) = - \frac2\pi \left(\chi(1) + \chi(3)\right) + O\left(\frac{k\_S}M \right).
}
The term $\chi(1) \chi(2) \chi(3)$ is here suppressed by an extra factor of $k\_S/M$, as follows from the discussion below eq.~\eqref{temp sum} in Appendix \ref{sec Wick}. Thus as long as $x$ is small enough, $X(x)$ can be obtained by summing over terms with $x - 1$ pairs of contracted fermions. For $x = 3$, this gives
\bel{
  X(3) = \frac4{\pi^2} \left(\frac 43 \chi(1) + \frac 89 \chi(3) + \frac 43 \chi(5)\right) + O\left(\frac{k\_S}M \right).
}
In general, $X(x)$ will be a weighted sum of $\chi(2x' - 1)$ for $x' \leq x$, with $\chi(1)$ and $\chi(2x - 1)$ having the highest weights. This is consistent with the fact that fermion fields and spin operators are not mutually local in the Ising CFT. It is remarkable that smoothing out converts the product of fields into a sum. This effect is similar to expanding a phase $\e^{\i \phi} \approx 1 + \i \phi$ in small fluctuations of the compact scalar $\phi$. Smoothing leads to an analogous expansion for Ising spins and fermions, whose $\Z_2$ target spaces would normally not support any ``small'' fluctuations.

Elementary methods are no longer sufficient when $x \gtrsim \log M$, as the number of contractions increases factorially with $x$. In these situations the sum over contractions is best handled by expressing it as the determinant of a Toeplitz matrix \cite{Montroll:1963wu, Schultz:1964fv, McCoy:2014}. While this method can be applied to calculating $X(x)$, it will fail at $x \sim M$ for reasons given in the main text.

Working with Toeplitz matrices really shines when applied to calculating the OPE $X_x \times X_y$, whose evaluation requires the smoothing of
\bel{\label{XxXy}
  X_x X_y = (-\i)^{y - x} \chi_{2x} \chi_{2x + 1} \cdots \chi_{2y - 1}.
}
The leading term in this smoothing is the sum $\Sigma_0$ over all possible sets of $r \equiv y - x$ contractions. It is convenient to express this sum as
\bel{
  \Sigma_0 \equiv \sum_{\sigma\in S_r} (-1)^\sigma \prod_{i = 0}^{r-1} (-\i) \wick[offset = 1.6ex]{\c \chi_{2x + 2i} \c \chi_{2x + 2\sigma_i + 1}},
}
where $S_r$ denotes the set of permutations of $\{0, 1, \ldots, r - 1\}$, and $(-1)^\sigma$ is the usual sign of the permutation. (Recall that, by eq.~\eqref{propagator Wick}, only contractions between sites of different parity are nonzero.) By further using eq.~\eqref{propagator Wick}, assuming $r \ll M/k\_S$, gives $\Sigma_0 = \det A$, with
\bel{\label{A}
  A_{ij} \equiv \frac2\pi \frac1{2i - 2j - 1}, \quad 0 \leq i, j \leq r - 1.
}
There exist powerful theorems to evaluate such determinants \cite{Szego:1952} (see \cite{Deift:2012} for a modern overview with historical comments and references to other seminal papers). In this case a simple numerical estimate will suffice, namely
\bel{
  \det A \approx (-1)^r \frac {c_0} {r^\delta},
}
with the limiting values $\delta = 0.2499(1)$ and $c_0 = 0.6450(1)$ obtained for matrix sizes up to $r = 3000$. The constant $c_0$, here obtained numerically, can be analytically calculated using Toeplitz methods and shown to be \cite{Wu:1966}
\bel{
  c_0 = \frac{\e^{1/4}\, 2^{1/12}}{{\tt A}^3},
}
where ${\tt A} \equiv \e^{1/12 - \zeta'(-1)} \approx 1.282$ is Glaisher's constant \cite{Deift:2012}. This means that, to leading order, the OPE of $X_x$ fields is
\bel{
  X_x \times X_y = (-1)^{x - y} \frac{c_0}{|x - y|^{1/4}} + O\left(\frac{k\_S |x - y|}M \right).
}


It is also possible to calculate the leading terms in this OPE that are different from the identity operator. These are obtained by summing over all possible terms in eq.~\eqref{XxXy} with $r - 1$ contractions. Each such term has two fermions, $\chi_{2x + 2x\_e}$ and $\chi_{2x + 2x\_o + 1}$, that are left uncontracted and instead individually get smoothed. The sum over these terms can be written as
\bel{\label{Sigma 2}
  \Sigma_2 \equiv -\i \!\! \sum_{x\_e,\, x\_o = 0}^{r - 1} \!\! (-1)^{x\_o - x\_e} \, \chi(2x + 2x\_e) \chi(2x + 2x\_o + 1) \det A^{(x\_e,\, x\_o)}.
}
Here $\det A^{(x\_e,\, x\_o)}$ is a minor of the matrix $A$ from eq.~\eqref{A}. A well known fact about minors is that they can be used to build the inverse of a matrix via
\bel{
  (A^{-1})_{x\_o,\, x\_e} = (-1)^{x\_e + x\_o} \frac{\det A^{(x\_e,\, x\_o)}}{\det A}.
}
While there exist theorems for asymptotics of Toeplitz minors (see \cite{Dehaye:2011} and references therein), these methods are not applicable to the matrix $A$ \cite{Kozlowski:2013}, and so in principle we have no immediate analytic handle on this computation. On the other hand, using the inverse matrix as above allows for an efficient numerical estimation of the needed prefactors. To leading order in $k\_S|x - y|/M$, all the fermion bilinears in \eqref{Sigma 2} can be approximated as
\bel{
  \chi(2x + 2x\_e) \chi(2x + 2x\_o + 1) \approx \frac{\i}2 E(x),
}
giving
\bel{
  \Sigma_2 = \frac{E(x)}2 \det A  \!\! \sum_{x\_e,\, x\_o = 0}^{r - 1} \!\! (A^{-1})_{x\_o,\, x\_e}.
}
The sum is easily evaluated numerically, giving a linear function $c_2' r$ with $c_2' \approx -\pi/2$, calculated for values of $r$ up to 5000. It is remarkable that this sum gives a linear function on the nose.

Putting everything together, the desired OPE can be expressed as
\bel{
  X_x \times X_y = \frac{(-1)^{x - y} c_0}{|x - y|^{1/4}} \left(\1 - \frac \pi4 |x - y| E(x) + \ldots \right) + O\left(\frac{k\_S |x- y|}{M} \right),
}
with the ellipses denoting terms with four and more smoothed fermions. These terms are guaranteed to be of higher order in $k\_S |x - y| / M$ than the fermion bilinears shown in the formula, and they can be ignored at this level of precision. Finally, replacing $\eps(x) \equiv \frac\pi2 E(x)$ gives the advertised result \eqref{OPE XX}.

\bibliographystyle{ssg}
\bibliography{IsingRefs}

\begingroup\raggedright\begin{thebibliography}{10}

\bibitem{Haag:1964}
R.~Haag and D.~Kastler, ``An Algebraic Approach to Quantum Field Theory,'' {\em
  Journal of Mathematical Physics} {\bf 5} (1964), no.~7 848--861.

\bibitem{GellMann:1962xb}
M.~Gell-Mann, ``{Symmetries of baryons and mesons},'' {\em Phys. Rev.} {\bf
  125} (1962) 1067--1084.

\bibitem{Wilson:1969zs}
K.~G. Wilson, ``{Nonlagrangian models of current algebra},'' {\em Phys. Rev.}
  {\bf 179} (1969) 1499--1512.

\bibitem{Mack:1969rr}
G.~Mack and A.~Salam, ``{Finite component field representations of the
  conformal group},'' {\em Annals Phys.} {\bf 53} (1969) 174--202.

\bibitem{Polyakov:1974gs}
A.~M. Polyakov, ``{Nonhamiltonian approach to conformal quantum field
  theory},'' {\em Zh. Eksp. Teor. Fiz.} {\bf 66} (1974) 23--42.

\bibitem{Rattazzi:2008pe}
R.~Rattazzi, V.~S. Rychkov, E.~Tonni, and A.~Vichi, ``{Bounding scalar operator
  dimensions in 4D CFT},'' {\em JHEP} {\bf 12} (2008) 031,
  \href{https://arxiv.org/abs/0807.0004}{{\tt 0807.0004}}.

\bibitem{ElShowk:2012ht}
S.~El-Showk, M.~F. Paulos, D.~Poland, S.~Rychkov, D.~Simmons-Duffin, and
  A.~Vichi, ``{Solving the 3D Ising Model with the Conformal Bootstrap},'' {\em
  Phys. Rev.} {\bf D86} (2012) 025022,
  \href{https://arxiv.org/abs/1203.6064}{{\tt 1203.6064}}.

\bibitem{Moore:1988qv}
G.~W. Moore and N.~Seiberg, ``{Classical and Quantum Conformal Field Theory},''
  {\em Commun. Math. Phys.} {\bf 123} (1989) 177.

\bibitem{Moore:1989yh}
G.~W. Moore and N.~Seiberg, ``{Taming the Conformal Zoo},'' {\em Phys. Lett.}
  {\bf B220} (1989) 422--430.

\bibitem{Ohya:2004}
M.~Ohya and D.~Petz, {\em Quantum entropy and its use}.
\newblock Springer Science \& Business Media, 2004.

\bibitem{Somma:2004}
R.~Somma, G.~Ortiz, H.~Barnum, E.~Knill, and L.~Viola, ``Nature and measure of
  entanglement in quantum phase transitions,'' {\em Phys.\ Rev.} {\bf A70}
  (2004), no.~4.

\bibitem{Casini:2013rba}
H.~Casini, M.~Huerta, and J.~A. Rosabal, ``{Remarks on entanglement entropy for
  gauge fields},'' {\em Phys. Rev.} {\bf D89} (2014), no.~8 085012,
  \href{https://arxiv.org/abs/1312.1183}{{\tt 1312.1183}}.

\bibitem{Lin:2018bud}
J.~Lin and {\DJ}.~Radi{\v c}evi{\' c}, ``{Comments on Defining Entanglement
  Entropy},'' \href{https://arxiv.org/abs/1808.05939}{{\tt 1808.05939}}.

\bibitem{Murray:1936}
F.~J. Murray and J.~v.~Neumann, ``On Rings of Operators,'' {\em Annals of
  Mathematics} {\bf 37} (1936), no.~1 116--229.

\bibitem{Halvorson2006}
H.~Halvorson and M.~Mueger, ``Algebraic Quantum Field Theory,''
  \href{https://arxiv.org/abs/math-ph/0602036}{{\tt math-ph/0602036}}.

\bibitem{Cardy:1996xt}
J.~L. Cardy, {\em {Scaling and renormalization in statistical physics}}.
\newblock Cambridge University Press, 1996.

\bibitem{Kadanoff:1970kz}
L.~P. Kadanoff and H.~Ceva, ``{Determination of an operator algebra for the
  two-dimensional Ising model},'' {\em Phys. Rev.} {\bf B3} (1971) 3918--3938.

\bibitem{Koo:1993wz}
W.~M. Koo and H.~Saleur, ``{Representations of the Virasoro algebra from
  lattice models},'' {\em Nucl. Phys.} {\bf B426} (1994) 459--504,
  \href{https://arxiv.org/abs/hep-th/9312156}{{\tt hep-th/9312156}}.

\bibitem{Milsted:2017csn}
A.~Milsted and G.~Vidal, ``{Extraction of conformal data in critical quantum
  spin chains using the Koo-Saleur formula},'' {\em Phys. Rev.} {\bf B96}
  (2017), no.~24 245105, \href{https://arxiv.org/abs/1706.01436}{{\tt
  1706.01436}}.

\bibitem{Jones:2019}
N.~G. Jones and R.~Verresen, ``Asymptotic Correlations in Gapped and Critical
  Topological Phases of 1D Quantum Systems,'' {\em Journal of Statistical
  Physics} {\bf 175} (2019), no.~6 1164–1213,
  \href{https://arxiv.org/abs/1805.06904}{{\tt 1805.06904}}.

\bibitem{Zou:2019dnc}
Y.~Zou, A.~Milsted, and G.~Vidal, ``{Conformal fields and operator product
  expansion in critical quantum spin chains},''
  \href{https://arxiv.org/abs/1901.06439}{{\tt 1901.06439}}.

\bibitem{Mong:2014ova}
R.~S.~K. Mong, D.~J. Clarke, J.~Alicea, N.~H. Lindner, and P.~Fendley,
  ``{Parafermionic conformal field theory on the lattice},'' {\em J. Phys.}
  {\bf A47} (2014), no.~45 452001, \href{https://arxiv.org/abs/1406.0846}{{\tt
  1406.0846}}.

\bibitem{OBrien:2017wmx}
E.~O'Brien and P.~Fendley, ``{Lattice supersymmetry and order-disorder
  coexistence in the tricritical Ising model},'' {\em Phys. Rev. Lett.} {\bf
  120} (2018), no.~20 206403, \href{https://arxiv.org/abs/1712.06662}{{\tt
  1712.06662}}.

\bibitem{Zou:2019iwr}
Y.~Zou and G.~Vidal, ``{Emergence of conformal symmetry in quantum spin chains:
  anti-periodic boundary conditions and supersymmetry},''
  \href{https://arxiv.org/abs/1907.10704}{{\tt 1907.10704}}.

\bibitem{Kitaev:2006lla}
A.~Kitaev, ``{Anyons in an exactly solved model and beyond},'' {\em Annals
  Phys.} {\bf 321} (2006), no.~1 2--111,
  \href{https://arxiv.org/abs/cond-mat/0506438}{{\tt cond-mat/0506438}}.

\bibitem{Radicevic:2019jfe}
{\DJ}.~Radi{\v c}evi{\'c}, ``{Abelian Bosonization, OPEs, and the `String
  Scale' of Fermion Fields},'' \href{https://arxiv.org/abs/1912.01022}{{\tt
  1912.01022}}.

\bibitem{Schultz:1964fv}
T.~D. Schultz, D.~C. Mattis, and E.~H. Lieb, ``{Two-dimensional Ising model as
  a soluble problem of many fermions},'' {\em Rev. Mod. Phys.} {\bf 36} (1964)
  856--871.

\bibitem{Kogut:1974ag}
J.~B. Kogut and L.~Susskind, ``{Hamiltonian Formulation of Wilson's Lattice
  Gauge Theories},'' {\em Phys. Rev.} {\bf D11} (1975) 395--408.

\bibitem{Susskind:1976jm}
L.~Susskind, ``{Lattice Fermions},'' {\em Phys. Rev.} {\bf D16} (1977)
  3031--3039.

\bibitem{DiFrancesco:1997nk}
P.~Di~Francesco, P.~Mathieu, and D.~Senechal, {\em {Conformal Field Theory}}.
\newblock Graduate Texts in Contemporary Physics. Springer-Verlag, New York,
  1997.

\bibitem{Weinberg:1996kr}
S.~Weinberg, {\em {The quantum theory of fields. Vol. 2: Modern applications}}.
\newblock Cambridge University Press, 2013.

\bibitem{Montroll:1963wu}
E.~W. Montroll, R.~B. Potts, and J.~C. Ward, ``{Correlations and spontaneous
  magnetization of the two-dimensional Ising model},'' {\em J. Math. Phys.}
  {\bf 4} (1963) 308--322.

\bibitem{McCoy:2014}
B.~M. McCoy and T.~T. Wu, {\em The two-dimensional Ising model}.
\newblock Courier Corporation, 2014.

\bibitem{Szego:1952}
G.~Szeg{\H o}, ``On certain Hermitian forms associated with the Fourier series
  of a positive function,'' {\em Comm. S{\'e}m. Math. Univ. Lund [Medd. Lunds
  Univ. Mat. Sem.]} {\bf 1952} (1952), no.~Tome Suppl{\'e}mentaire 228--238.

\bibitem{Deift:2012}
P.~Deift, A.~Its, and I.~Krasovsky, ``Toeplitz matrices and Toeplitz
  determinants under the impetus of the Ising model. Some history and some
  recent results,'' \href{https://arxiv.org/abs/1207.4990}{{\tt 1207.4990}}.

\bibitem{Wu:1966}
T.~T. {Wu}, ``{Theory of Toeplitz Determinants and the Spin Correlations of the
  Two-Dimensional Ising Model. I},'' {\em Physical Review} {\bf 149} (1966),
  no.~1 380--401.

\bibitem{Dehaye:2011}
P.-O. Dehaye, ``On an Identity due to Bump and Diaconis, and Tracy and Widom,''
  {\em Canadian Mathematical Bulletin} {\bf 54} (2011), no.~2 255--269.

\bibitem{Kozlowski:2013}
K.~K. Kozlowski, ``On lacunary Toeplitz determinants,''
  \href{https://arxiv.org/abs/1310.2584}{{\tt 1310.2584}}.

\end{thebibliography}\endgroup

\end{document}